\documentclass[aps,prb,reprint,superscriptaddress]{revtex4-2}
\usepackage{graphicx}
\usepackage{amsmath}
\usepackage{braket}
\usepackage{bm}
\usepackage{color} 
\usepackage{booktabs}

\newcommand{\uspace}{\mskip3mu}

\begin{document}
\title{High-throughput study of the anomalous Hall effect} 

\author{Jakub \v{Z}elezn\'{y}}
\affiliation{Institute of Physics, Czech Academy of Sciences, Cukrovarnick\'{a} 10, 162 00 Praha 6 Czech Republic}
\author{Yuta Yahagi}
\affiliation{Department of Applied Physics, Tohoku University, Sendai, Miyagi, Japan}
\author{Carl\'es-Gomez Ollivella}
\affiliation{Institute of Physics, Czech Academy of Sciences, Cukrovarnick\'{a} 10, 162 00 Praha 6 Czech Republic}
\author{Yang Zhang}
\affiliation{Department of Physics, Massachusetts Institute of Technology, Cambridge, Massachusetts 02139, USA}
\author{Yan Sun}
\affiliation{Shenyang National Laboratory for Materials Science, Institute of Metal Research, Chinese Academy of Sciences, Shenyang 110016, China}

\begin{abstract}
Despite being known for a long time the anomalous Hall effect still attracts attention because of its complex origins, its connection to topology and because it serves as a useful probe of the magnetic order. Here we study the anomalous Hall effect using automatic high-throughput calculation scheme. We calculate the intrinsic anomalous Hall effect in 2871 ferromagnetic materials. We use these results to study general properties of the anomalous Hall effect such as its dependence on the strength of the spin-orbit coupling or magnetization. We also examine the origin of the anomalous Hall effect in the materials with the largest effect and show that the origin of the large anomalous Hall effect is usually associated with symmetry protected band degeneracies in the non-relativistic electronic structure, typically mirror symmetry protected nodal lines. Additionally, we study the dependence of the anomalous Hall effect on the magnetization direction, showing that in many materials it differs significantly from the commonly assumed expression $\mathbf{j}^\text{AHE} \sim \mathbf{M} \times \mathbf{E}$.
\end{abstract}

\maketitle

\section{Introduction}
When magnetic field is applied to a metal in which current is flowing, a transverse electrical current appears. This is the so-called classical Hall effect. In some magnetic materials a transverse current appears even in absence of external magnetic field, an effect known as the anomalous Hall effect (AHE) \cite{Nagaosa2010}. It originates from the time-reversal symmetry breaking due to the magnetic order and the relativistic spin-orbit coupling (although in non-collinear magnetic materials it can also exist in absence of spin-orbit coupling). Historically, it has been mainly studied in ferromagnetic materials, however, it has recently been shown that it can also exist in some antiferromagnetic materials \cite{Chen2014,Kubler2014,Nakatsuji2015,Nayak2016,Smejkal2020,Feng2020observation,Smejkal2021}.

Although the anomalous behavior of the Hall effect in ferromagnetic materials has already been noticed by Hall in 1881 \cite{Hall1881}, the AHE is still an actively investigated effect. Theoretically, it is known that its origin can be split into two main categories: the intrinsic and extrinsic. The intrinsic contribution is determined purely by the electronic structure of a perfect crystal. In contrast, the extrinsic contributions originate from electron scattering. Here, we will only consider the intrinsic contribution as it is much easier to calculate and has a universal value for each material, whereas the extrinsic contribution will dependent on presence of impurities and experimental conditions such as temperature. A remarkable aspect of the intrinsic contribution is that it describes a non-dissipative transport. Furthermore, it can be understood in a geometrical sense: the intrinsic contribution is given by integral of the Berry curvature of each occupied band. In this way it is also connected to topology since in insulators the integral of the Berry curvature determines the Chern number topological invariant. In metallic materials, dissipation cannot be avoided since a dissipative current will always be present together with the AHE. However, in topologically non-trivial magnetic insulators, i.e. insulators with nonzero Chern number, only the intrinsic AHE is present (known as the quantum AHE \cite{Chang2022}), which then allows for a truly dissipationless transport at the edge.

The intrinsic AHE can be in many materials described well by \emph{ab-initio} calculations based on density functional theory (DFT). Here, we utilize automatic DFT calculations to calculate the intrinsic AHE in a large number of magnetic materials. Such high-throughput computational approach has seen intense development in recent years and has recently been also applied to transport effects \cite{Zhang2021,Sakai2020,Noky2020}. Our work provides a large reference database of the intrinsic AHE values and furthermore, we use the large number of calculated materials to explore general properties of the AHE.

We mainly focus on ferromagnetic materials here. We only consider collinear magnetic materials since our calculation procedure is not applicable to non-collinear magnetic materials. Unfortunately, no comprehensive experimental database of ferromagnetic materials exist. MAGNDATA \cite{Gallego2016a,Gallego2016b}, the only large experimental database of magnetic materials, contains mostly antiferromagnets . Because of this we use as a source of materials, the computational materials database Materials Project \cite{Jain2013}. This database contains both real crystal structures and materials computationally predicted to be potentially stable, however, it includes no information about experimental magnetic order since that is not available for most materials. Instead, a ferromagnetic configuration is used as a starting point for the DFT calculation. If this converges to ferromagnetic state this signifies that the material is likely magnetic, although its real magnetic order could be different. Here we consider the ferromagnetic order for the magnetic materials from the Materials Project, since theoretically determining the ground state magnetic order is a very complex problem, beyond the scope of this project. This approach means that the calculated AHE value for any given material can not be used without further verification of the magnetic order, however, importantly even in cases where the actual magnetic order differs from the assumed ferromagnetic one, the calculation can be used to infer general, statistical properties of the AHE. A similar computational approach has been recently used for a high-throughput calculations of the anomalous Nernst effect.\cite{Sakai2020}.

We explore the statistical properties of the AHE. In particular we focus on the relation between the AHE magnitude, spin-orbit coupling strength, magnetization or symmetry. Although, a relation between the magnitude of magnetization and spin-orbit coupling exist, we find that this is relatively weak and large AHE can exist also in materials with relatively small magnetization and light elements. Analogously to our previous study of the spin Hall effect (SHE), we find that large AHE is typically associated with symmetry protected degeneracies in the non-relativistic electronic structure \cite{Zhang2021}. These are typically mirror symmetry protected nodal lines. They are split by the spin-orbit coupling interaction and the two slit bands can then carry large Berry curvature, leading to large AHE, if they appear at the Fermi level. Consequently, we find a statistical link between the symmetry and the AHE magnitude, although this link is fairly weak and the most apparent for large AHE materials.

Additionally, we study in a subset of the materials, the dependence of the AHE on the rotation of the magnetization. We find, that in contrast to usual expectations, the magnetization dependence can deviate significantly from the simple relation $\mathbf{j}^\text{AHE} \sim \mathbf{M} \times \mathbf{E}$, where $\mathbf{M}$ is the magnetization, $\mathbf{E}$ is the electric field and $\mathbf{j}^\text{AHE}$ is the AHE current.  This formula has to be satisfied in a fully isotropic materials (such as polycrystals) but does not in general hold in crystalline materials. We find that this relation approximately holds in some high-symmetry materials, however, even in high-symmetry materials significant deviations are typically present and in some materials the AHE dependence can be completely different. In low symmetry materials, the AHE dependence cannot usually be described by the simple formula at all. The deviation of AHE from the simple relation has been studied previously \cite{HongbinZhang2011}, however, here we show that the deviations are very common and without further verification the simple relation cannot be used in single crystalline materials at all.

Most of the materials we have calculated are metallic, however, our calculations also include several hundred insulators. Our results suggest that all of these are topologically trivial and thus do not exhibit the quantum anomalous Hall effect. Similarly to the case of the SHE, our results show that large AHE values are very rare. None of the calculated materials has much larger AHE magnitude than the well known case of Ni, thus suggesting existence of a practical limit to the intrinsic AHE magnitude.

In addition to the AHE we also calculate the ordinary conductivity tensor, using the constant relaxation time approximation. Since the relaxation time is unknown and sample and conditions dependent, these calculations are less general than the AHE calculations. Nevertheless, they can be used as a rough estimate of the conductivity magnitude or broad trends and we use them to obtain the anomalous Hall angle: the ratio between the anomalous Hall and longitudinal conductivities.

\section{Workflow}

We consider materials that were found to be magnetic in the Materials Project database \cite{Jain2013}. Vast majority of these materials assume ferromagnetic order, although a few are antiferromagnetic. This magnetic order is not necessarily the proper ground state magnetic order, however, we assume it here for simplicity. We exclude materials for which LDA+U approach was used in the Materials Project. We have calculated 2871 materials in total. We use Aiida \cite{UHRIN2021110086,Huber2020} for automatization of the calculations.

For each material, we use a three step calculation procedure, analogous to the one used in our high-throughput study of the spin Hall effect \cite{Zhang2021}. First, we obtain a ground state electronic structure using the FPLO DFT code \cite{FPLO}. Afterwards, we construct a Wannier tight-binding like Hamiltonian. This procedure is straighforward in FPLO since FPLO uses a local basis set. We use the same set for the construction of the Wannier Hamiltonian. The Hamiltonian thus differs from the full DFT Hamiltonian only in that matrix elements that are small are set to zero. Consequently, the Wannier Hamiltonian is very accurate and can easily be constructed automatically, which is a significant advantage over the often used approach based on the maximally localized Wannier functions \cite{Mostofi2014}. In the last step, we use the Wannier Hamiltonian for a linear response conductivity calculation.

Although our main aim is the intrinsic AHE, we include disorder description into the linear response calculation, via the so-called constant $\Gamma$ approximation. $\Gamma$ is a parameter that describes the disorder, corresponding to a constant relaxation time $\tau = \hbar/2\Gamma$. We use the linear response formulas derived in Ref. \cite{freimuth2014}.

\section{Results}

The full database of all our calculations is available at \cite{HTP_heroku} and in the Supplementary Material. Here we discuss general statistical properties of the AHE and its dependence on the magnetization direction. In addition, we discuss the origin of AHE in the materials with largest AHE and its relation to nodal lines in the non-relativistic electronic structure.

\subsection{Statistics}

\begin{figure*}
    \includegraphics[width=\textwidth]{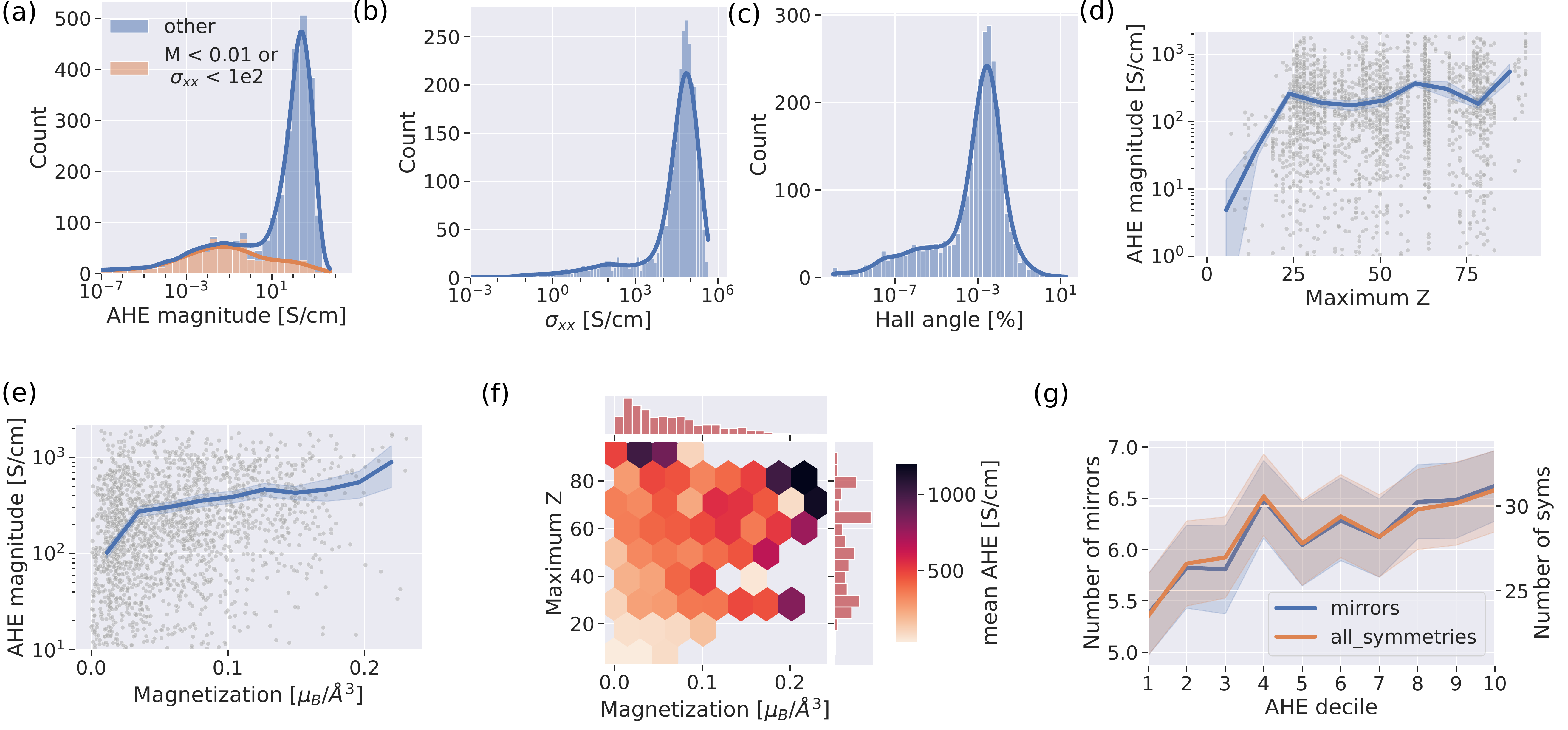}%
    \caption{\label{fig:statistics} \textbf{AHE statistics}. In the histogram plots we show the kernel density estimation along with the histogram.
    (a) The histogram of all the calculated AHE values for $\Gamma = 0.1 \uspace \text{meV}$. We separate materials with either small magnetization density (less than $0.01 \uspace \mu_B/\AA^3$) or very low conductivity (less than $100 \uspace \text{S/cm}$).
    (b) The histogram  of the $\sigma_{xx}$ conductivities for $\Gamma = 10 \uspace \text{meV}$.
    (c) The histogram of the Hall angle calculated using $\Gamma = 0.1 \uspace \text{meV}$ for AHE and $\Gamma = 10 \uspace \text{meV}$ for the conductivity. 
    (d) The dependence of the AHE magnitude on the maximum atomic number of each compound. The gray dots show the individual calculations. The blue line is obtained by separating all maximum atomic number values into 10 bins and averaging within each bin. The light blue region shows the 95\% confidence intervals estimated using the empirical bootstrap method.
    (e) Same as (d) but for magnetization density, instead of the maximum atomic number.
    (f) Each hexagonal bin in this plot shows the mean value of AHE for materials with corresponding maximum atomic number and magnetization density. We also show histograms of the magnetization density values and maximum atomic numbers.
    (g) Average number of mirrors (left) and total symmetry operations (right) as a function of the AHE decile for materials with AHE magnitude larger than $10 \uspace \text{S/cm}$. This means that we separate all AHE values by their magnitude into 10 bins such that each bin has the same number of values and then calculate the average number of mirrors resp. symmetry operations in each bin. Thus, decile 1 corresponds to 10\% materials with the smallest AHE and decile 10 to the largest 10\%.
    }
\end{figure*}

In Fig. \ref{fig:statistics}(a) we give the histogram of the AHE magnitudes for all calculated materials. This shows that materials with large AHE (larger than $\sim 1000 \uspace \text{S/cm}$) are very rare. The tail of the histogram below around $\sim 1 \uspace \text{S/cm}$ is dominantly due to materials with very low magnetization density (less than $0.01 \uspace \mu_B/\text{\AA}^3$) or very low conductivity (less than $100 \uspace \text{S/cm}$), which corresponds to insulators or semimetals with low density of states at the Fermi level. Most metals with non-negligible magnetization thus have AHE between $10 - 1000 \uspace \text{S/cm}$.

In Fig. \ref{fig:statistics}(b) we show the histogram of the $xx$ components of the conductivity tensor for all calculated materials. Since the value of the broadening parameter $\Gamma$ is unknown we set here $\Gamma=10\uspace \text{meV}$ for all materials. This allows comparison between materials or estimating general trends, however, it cannot be used as an accurate quantitative estimate of the conductivity in individual materials. The full $\Gamma$ dependence is given at \cite{HTP_heroku}. As can be seen in Fig. \ref{fig:statistics}(b) the conductivity of most metals lie between $10^4 - 10^6 \uspace \text{S/cm}$.

Using the conductivity, the (anomalous) Hall angle, i.e. the ratio $\sigma^{AHE}_{xy}/\sigma_{xx}$ can also be evaluated. Histogram of all the Hall angles is given in Fig. \ref{fig:statistics}(c). Interestingly, we find that the distribution of Hall angles drops much less slowly for large Hall angles than the AHE or the conductivity. This thus suggests that although it may be very hard to obtain very large intrinsic AHE, it may be more feasible to find materials with large Hall angles. The reason for this behavior is that the conductivity and the AHE are not significantly correlated in metallic materials and thus it can happen that a material has large AHE but low conductivity.

In Fig. \ref{fig:statistics}(d) we give the relation between the maximum atomic number of each compound and the magnitude of AHE. We use the maximum atomic number as an estimate of the strength of the spin-orbit coupling: materials with large atomic numbers have larger spin-orbit coupling. We find that AHE tends to increase with the atomic number, but the dependence is strong only for atoms with small atomic number. For atomic numbers beyond 25, the dependence becomes very weak. In Fig. \ref{fig:statistics}(e) we give an analogous plot for the relation between the magnetization density and the AHE magnitude. Here we observe a more clear trend, however, similarly as for the atomic number, the dependence of AHE on magnetization is only strong for small magnetizations. In Fig. \ref{fig:statistics}(f) we show a 2D histogram that illustrates the AHE dependence on both the magnetization and the atomic number. This shows that materials with large AHE are those with either large magnetization or large atomic number or both. We find that even low magnetization materials can have large AHE if they have large atomic number and vice versa.

In our high-throughput study of the SHE we have found that large SHE is typically associated with mirror symmetry protected nodal lines in the non-relativistic electronic structure. As discussed below, our analysis of the materials with largest AHE also reveals that the AHE hotspots in the materials with largest AHE are usually associated with symmetry protected band degeneries, most commonly mirror symmetry protected nodal lines. Consequently, we observe a statistical relationship between symmetry and AHE magnitude, see Fig. \ref{fig:statistics}(g). However, similarly to the case of the SHE \cite{Zhang2021}, this effect is fairly subtle. As shown in Fig. \ref{fig:statistics}(g), we observe a very similar relation between AHE magnitude and symmetry both when we consider number of mirror symmetry operations and total number of symmetries. Since the number of mirrors is strongly correlated with the number of total symmetries it is not possible to say whether it is mainly the mirrors that play a role, or all symmetry operations.
 
\subsection{Magnetization rotation}

For a subset of materials we rotate the magnetization to study the dependence of AHE on magnetization direction. We always rotate from the [010] direction to the [001] direction. Here the directions are given in a cartesian coordinate system. The coordinate system is always chosen such that [001] direction is along the $c$ axis of the conventional lattice and for all groups except triclinic the [010] direction is aong the $b$ axis (for triclinic $b$ is in the (100) plane). Since AHE is described by an antisymmetric 3x3 tensor, it can equivalently be described by a (pseudo)-vector: $\mathbf{j} = -\mathbf{h} \times \mathbf{E}$, where the $\mathbf{h}$ is the Hall vector defined as $\mathbf{h} = (\sigma^\text{AHE}_{yz}, -\sigma^\text{AHE}_{xz}, \sigma^\text{AHE}_{xy})$. Along high-symmetry directions, the Hall vector is constrained by symmetry to lie along the same axis as magnetization, however, in general their directions differ. Below we discuss general properties of the rotation dependence and give one material as example. The detailed results for all materials can be found in the Supplementary Material.

We separate the materials by symmetry into 3 types depending on the symmetry of the Hall vector along the magnetization rotation. For all the materials that we have considered the Hall vector is constrained to lie along the same axis for the [010] magnetization direction. In group $A$ the Hall vector in addition lies along the magnetization for the [110] and [001] magnetization and its magnitude is the same for the [010] and [001] directions. In group $B$, the Hall vector is  constrained to lie along magnetization for the [001] directions, but not the [110] direction and the magnitudes for [010] and [001] directions are not the same. In group $C$ the Hall vector is furthermore not constrained to lie in the along the magnetization even for the [001] direction.

\begin{figure}
    \includegraphics[width=\columnwidth]{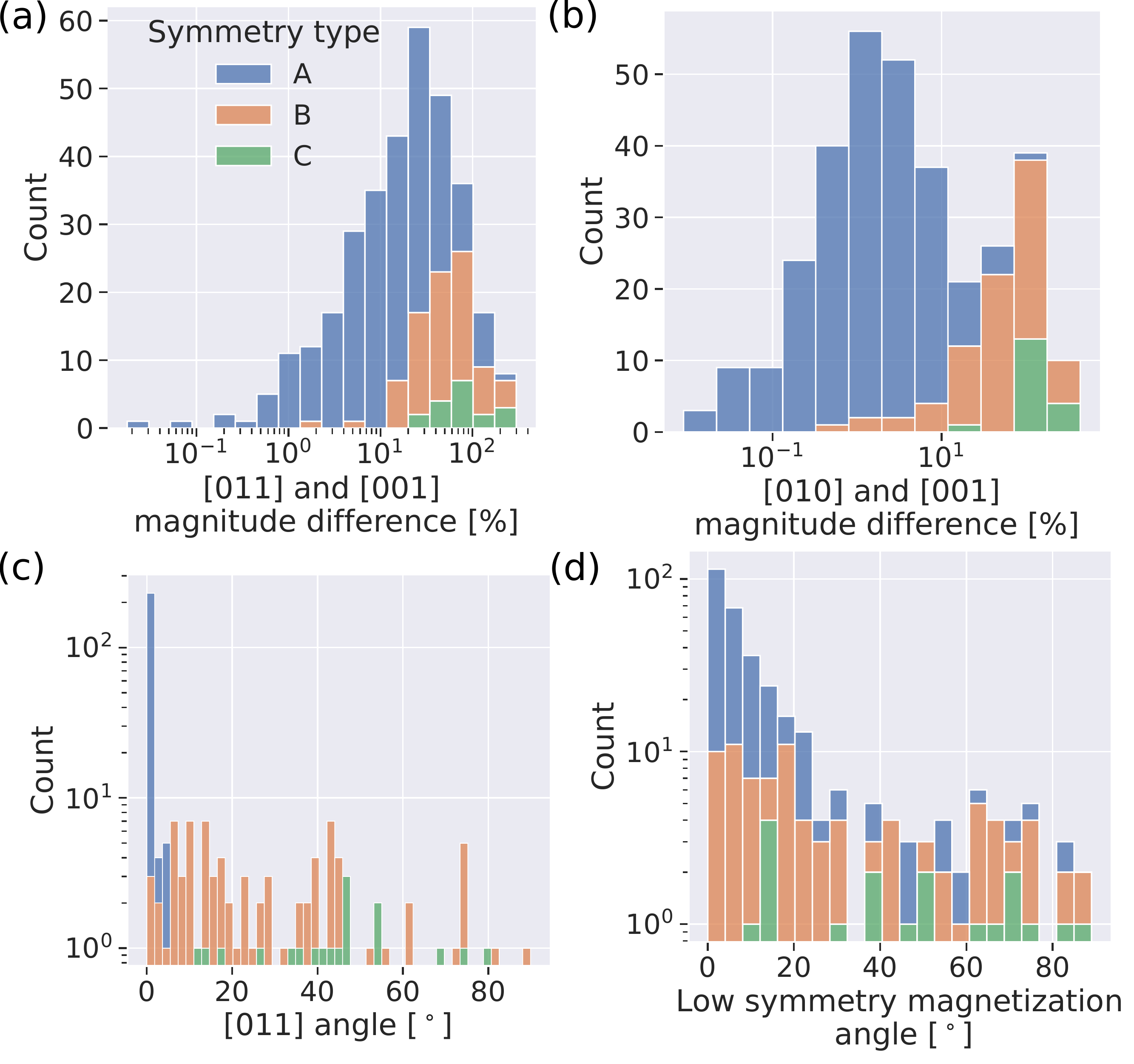}%
    \caption{\label{fig:rotations} \textbf{Magnetizatin rotation statistics}.
    (a,b) Histogram of the differences between Hall vector magnitudes for magnetizations along (a) [011] and [001] directions and (b) [010] and [001] directions. For materials with symmetry type A, the magnitudes are constrained to be the same along the [010] and [001] directions. In practice we find that they usually differ by few \% or less, due to numerical inaccuracy.
    (c,d) The angle between magnetization and the Hall vector for magnetization along (c) [011] direction and (d) $[0,\cos(\pi/8),\sin(\pi/8)]$ direction, which is a low symmetry direction.
    }
    
\end{figure}

As shown in Fig. \ref{fig:rotations} we find that the magnitude of the Hall vector can strongly depend on the magnetization direction. This is more pronounced in the lower symmetry groups, but even in the highest symmetry material a significant differences between the Hall vector magnitudes for the [011] and [001] magnetization direction are commonly found (Fig. \ref{fig:rotations}(a)). For the difference between [010] and [001] magnetization directions, a large difference is also found for materials with symmetry type $B$ and $C$ (we also find a small difference for materials with symmetry type $A$, however, this is a numerical error). Furthermore, we find that the Hall vector can deviate significantly from the magnetization direction, if it is not constrained by symmetry (see Figs. \ref{fig:rotations}(c),(d). This is again more pronounced for the lower symmetry materials, but even in the highest symmetry materials significant deviations are often present outside of the [010], [011] and [001] magnetization directions. These calculations show that the assumption of $\mathbf{j}^{AHE} \sim \mathbf{M} \times \mathbf{E}$ can only be used in polycrystalline materials. In single crystals, significant deviations from this expression are common even for crystals with very high symmetry. We note that we typically see only a small changes in the magnitude of the magnetic moments as they are rotated, thus the anisotropy of the AHE are mainly due to AHE itself, not due to changes of magnetic moments.

\begin{figure}
    \includegraphics[width=\columnwidth]{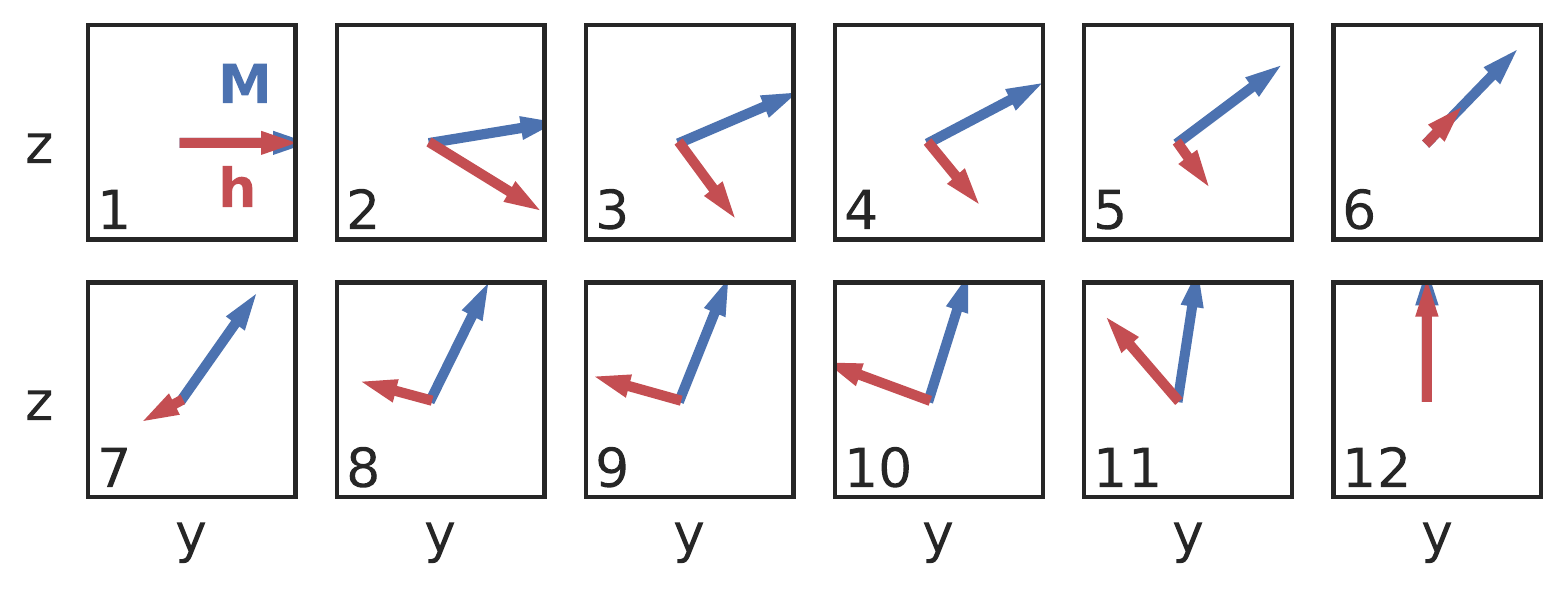}%
    \caption{\label{fig:ind_rotation} \textbf{Magnetization rotation for HfMnTl}. 
    The dependence of the Hall vector on the magnetization direction in HfMnTl. Here the magnetization is rotated from the [010] direction to the [001] direction. The blue arrows show the magnetization direction, the red arrows show the Hall vector, which has been scaled so that it's magnitude is comparable to the length of the magnetization arrow. The individual plots show different orientations with the number denoting the step along the rotation.}
    
\end{figure}

An example of a significant deviation from the simple dependence of the Hall vector on magnetization is given in Fig. \ref{fig:ind_rotation}, which shows the evolution of Hall vector for rotation of the magnetization in the $y-z$ plane for HfMnTl \cite{osti_1279070}. Even though this material has the highly symmetric $F\bar{4}3m$ space group, which constrains the Hall vector to lie along magnetization for [010], [011] and [001] magnetization directions, we find that as the magnetization is rotated from the [010] direction to [001] direction the Hall vector rotates in the opposite direction to the magnetization.

\subsection{Analysis of top materials}

In Ref. \cite{Zhang2021} it was found that in materials with large intrinsic SHE, the hotspots of the spin Berry curvature are typically associated with symmetry protected band degeneracies in the non-relativistic band structure. These degeneracies are split by spin-orbit coupling and the two bands then carry large spin Berry curvature if the Fermi level is positioned in between them. Usually these degeneracies are associated with mirror-symmetry protected nodal lines. 

To explore whether this also applies to the AHE, we have studied the Berry curvature hotpots in 6 materials with the largest AHE in our database. The materials we consider are U$_2$P$_2$, GdTmRh$_2$, Ni, Eu$_2$SeO$_2$, MnCoPt$_2$ and CeTe$_2$, as shown in Table \ref{table}. Note that we do not consider MnInRh$_2$, which is the material with largest AHE in the database since the convergence in this material is poor. We identify degeneracies in the non-relativistic band structure by finding points where bands are closer than $0.1\uspace \text{eV}$. In ferromagnetic materials crossings between bands with opposite spin are common in the non-relativistic band structure. These appear without any association with symmetry since in collinear magnetic system the opposite spins are not coupled without spin-orbit coupling. These crossings are split by SOC, however, we find that this does not usually result in a large Berry curvature. Thus we focus on crossings of bands with the same spin, which are usually protected by some symmetry.

\begin{table}
\begin{tabular}{llrr}
\toprule
formula &         id &  $||\mathbf{h}||$ [S/cm] &  spacegroup \\
\midrule
  U$_2$PN$_2$ &    mp-5381 &               3023.35 &         164 \\
GdTmRh$_2$ & mp-1184489 &               2848.17 &         225 \\
     Ni &      mp-23 &               2437.72 &         225 \\
Eu$_2$SeO$_2$ &  mp-753314 &               2428.01 &         139 \\
MnCoPt$_2$ & mp-1221704 &               2400.29 &         123 \\
  CeTe$_2$ &  mp-505536 &               2342.23 &         129 \\
\bottomrule
\end{tabular}
\caption{The six materials with the largest AHE (exluding MnInRh$_2$) for which we have studied origin of AHE in detail. Here \emph{id} refers to the Materials Project id, $||\mathbf{h}||$ is the Hall vector magnitude and \emph{spacegroup} is the non-magnetic spacegroup number.}
\label{table}
\end{table}

\begin{figure}
    \includegraphics[width=\columnwidth]{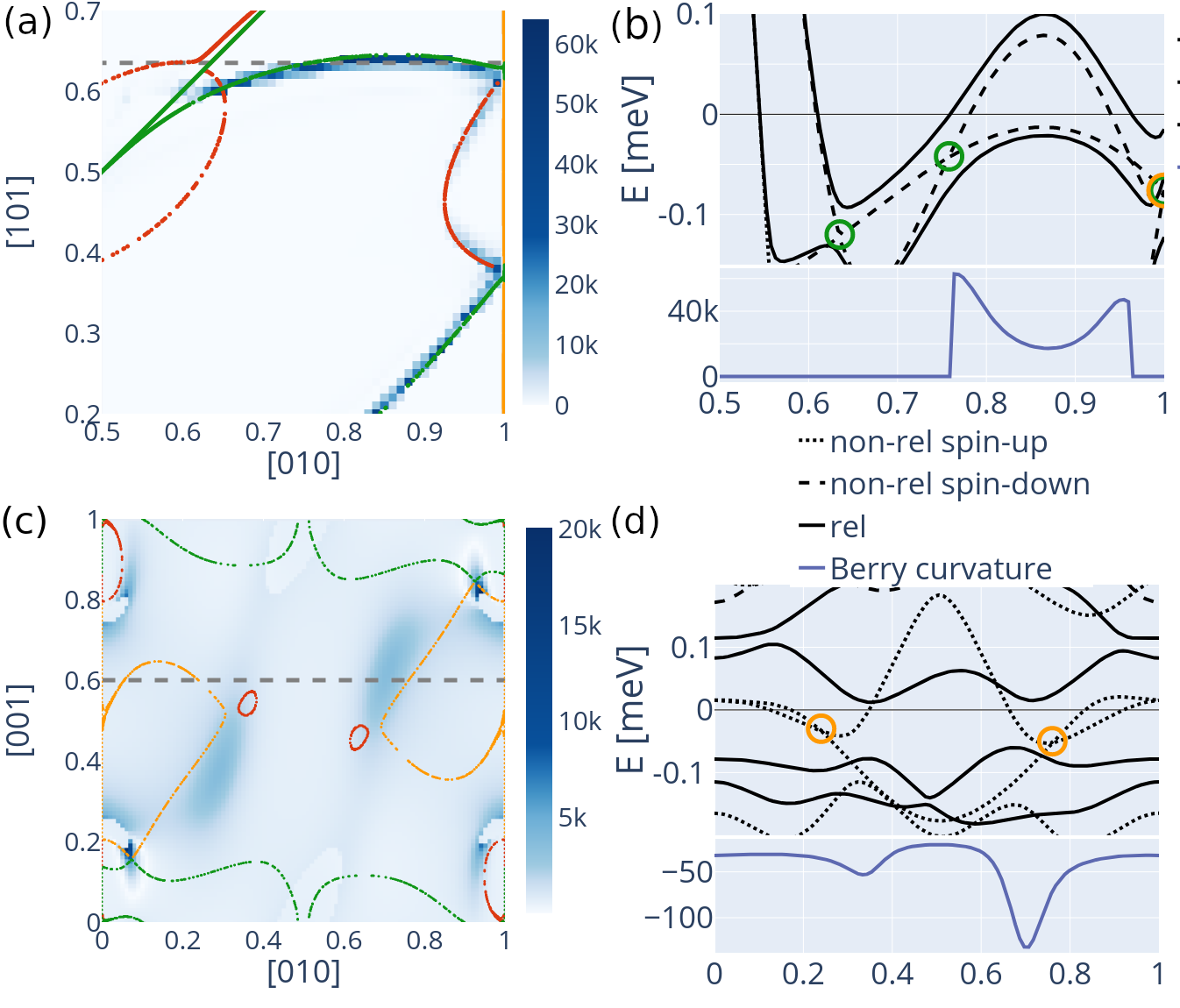}%
    \caption{\label{fig:nodal_lines} \textbf{Nodal lines in Ni and U$_2$PN$_2$}. In these plots we give the Berry curvature summed over all occupied bands, given in units of S/cm \AA$^{-3}$. When integrated over the first Brillouin zone, this directly gives the intrinsic AHC. (a) Berry curvature distribution and nodal lines in the (110) mirror plane in Ni. Only a part of the mirror plane containing the main hotspots is shown, to better highlight the overlap with nodal lines. The plot is given in relative coordinates of the [010] and [101] vectors of the primitive reciprocal lattice. (b) The non-relativistic and relativistic band structure of Ni and the associated Berry curvature along the dashed line denoted in (a). Here, the green and yellow circles denote the corresponding nodal lines in (a). (c,d) Sam as (a,b) but for the (100) mirror plane in U$_2$PN$_2$.
    }
    
\end{figure}

We give some of the results for U$_2$PN$_2$ and Ni in Fig. \ref{fig:nodal_lines}, the full results are available at \cite{HTP_heroku}. We find that in Ni, the large AHE comes mainly from sharp hotspots, which all very closely overlap with mirror symmetry protected nodal lines in the non-relativistic electronic structure. The main hotspots are located in 6 symmetry equivalent mirror planes (110), (101), (011), (-101), (01-1) and (1-10), which is shown in Fig. \ref{fig:nodal_lines}(a). In Fig. \ref{fig:nodal_lines}(b) we give the band structure along a path through this plane, which demonstrates how the non-relativistic crossing results in a large Berry curvature if the SOC split bands are at the Fermi level.

In U$_2$PN$_2$ we find that the main Berry curvature hotposts are located in 3 equivalent mirror planes (100), (010) and (110) and in the (1-10) plane, which is not a mirror plane, but contains a rotation symmetry protected nodal lines. In Fig. \ref{fig:nodal_lines}(c) we give the Berry curvature and nodal lines in the (100) mirror plane and in Fig. \ref{fig:nodal_lines}(d) we give a band structure on a path through this plane. We find that also in this case the hotspots are most likely associated with nodal lines, however, the overlap is less clear since the large SOC shifts the bands considerably from the non-relativistic bands. As shown in \cite{HTP_heroku} the hotspots within the (1-10) plane are associated with the rotation symmetry protected nodal lines.

As shown in \cite{HTP_heroku}, in all of the materials that we have studied, we find that the Berry curvature hotspots are mainly located along high-symmetry planes or lines, suggesting a connection to symmetry. In many cases we can identify non-relativistic nodal lines from which these hotspots originate, however, this is not always possible clearly since often the SOC is large and many bands are present at the Fermi level. This thus suggests that symmetry plays an important role for existence of large AHE and in particular that materials with high symmetry and especially many mirror planes are beneficial for large AHE.

\section{Methods}

\subsection{DFT}

\begin{figure}
    \includegraphics[width=\columnwidth]{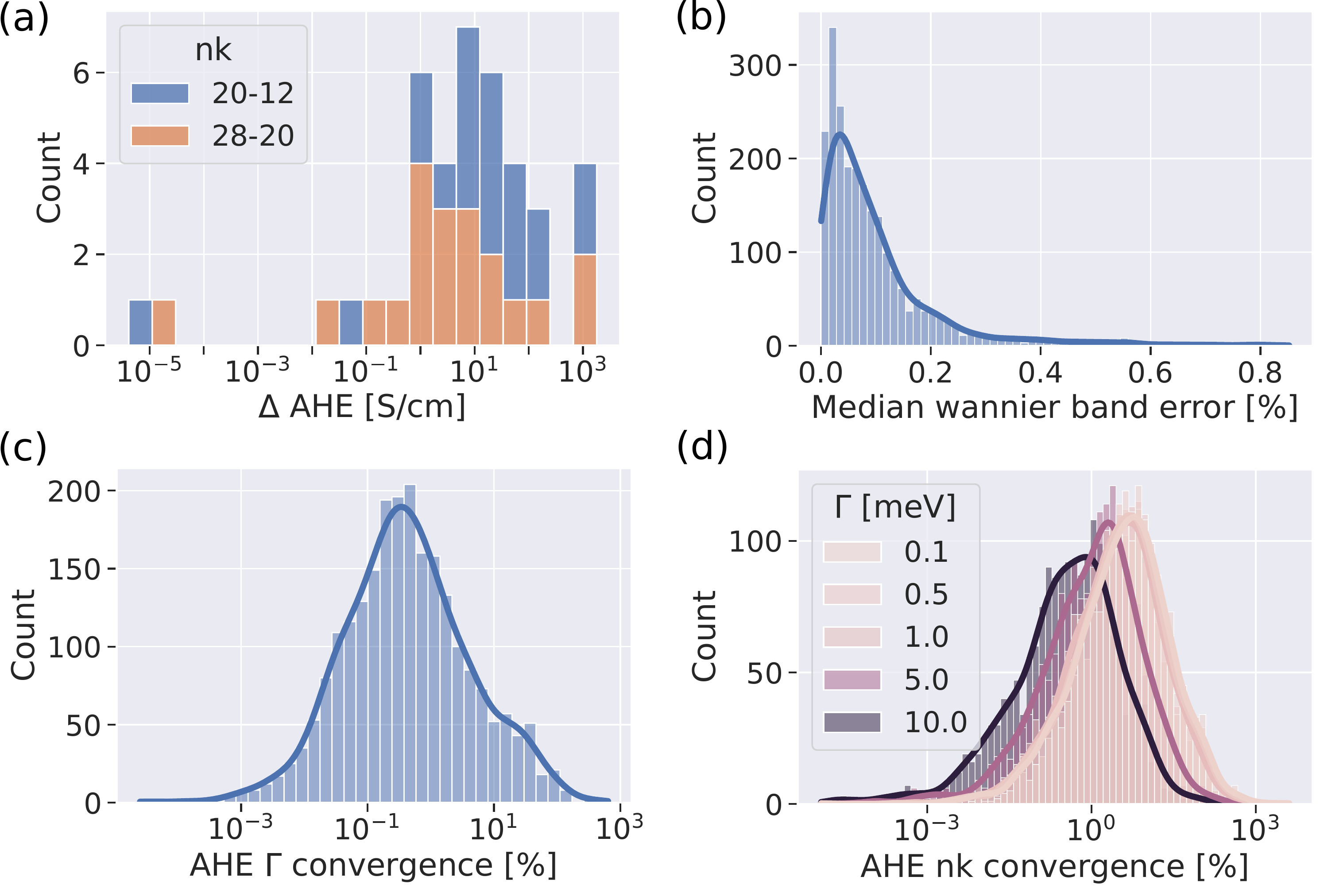}%
    \caption{\label{fig:convergence} \textbf{Convergence tests}
    (a) The dependence of the xy component of the AHE on the DFT k-mesh for 20 materials. The plot show the histogram of the differences between the 20x20x20 and 12x12x12 meshes and between 28x28x28 and 20x20x20 meshes.
    (b) The difference between the DFT bands and the bands obtained from the Wannier Hamiltonian for 2312 materials. We calculate the relative difference for every k-point and band. The plot shows the histogram of the median differences for each material.
    (c) Histogram of the $\Gamma$ convergence for materials with AHE larger than $0.1 \uspace \text{S/cm}$. The $\Gamma$ convergence is defined as the relative difference between the xy component of AHE for $\Gamma = 0.1\ \uspace \text{meV}$ and $\Gamma = 0.5 \uspace \text{meV}$.
    (d) Convergence in the number of k-points for the linear response calculation. Histogram of the relative difference of the $xy$ component of AHE between the 250x250x250 and 125x125x125 meshes for all materials with AHE larger than $10 \uspace S/cm$.
    }
\end{figure}

The FPLO DFT calculation utilize the PBE-GGA exchange correlation potential \cite{PBEGGA} and we use 20x20x20 k-point mesh for every material. We have tested the AHE dependence on the number of k-points for the DFT calculation for 20 materials with varying number of unit cell sizes. As shown in Fig. \ref{fig:convergence}(a) we find that the difference between 28x28x28 mesh and 20x20x20 is for most materials below 10 S/cm, although larger errors can also happen. 

Since FPLO does not include magnetic symmetry, we use the non-magnetic space group for the input, which is obtained by setting all magnetic atoms with different magnetic moments as chemically distinct atoms. Apart from the rotation dependence calculations, we always set the direction of magnetic moments along the $c$ axis of the conventional lattice.

For the construction of the Wannier Hamiltonian we use the full DFT basis set, apart from the core states. We use no energy windows. We set to zero all the matrix elements smaller than $0.001 \uspace \text{eV}$ or those involving states that are more than $25 \uspace  a_0$, where $a_0$ is the Bohr radius. We have tested the AHE calculation for BCC Fe with the energy cutoff set to $0.0001 \uspace \text{eV}$ and the length cutoff to $50 \uspace  a_0$ and have found that the difference is well below $1\%$ for all considered $\Gamma$ values. The accuracy of the Wannier Hamiltonian can be further tested by comparing the band structure obtained from the DFT calculation and the band structure from the Wannier Hamiltonian, which we have calculated for most (although not all) materials. As shown in Fig. \ref{fig:convergence}(b) we find a very good agreement between the bandstructures in most materials. 

\subsection{Linear response}
 To obtain the Kubo formula for conductivity from the formulas given in Ref. \cite{freimuth2014}  we replace the torque operator $-{\cal T}_i$ by current density operator $-e \hat{v}_i  / V$, where $e$ is the elementary charge ($e > 0$), $\hat{v}_i$ is the $i$-th component of the velocity operator and $V$ is the unit cell volume. Then the following formulas are obtained

\begin{align}
	\sigma^{{\rm even}}_{ij} = &
	\frac{e^2\hbar}{V\pi} \sum_{\textbf{k}nm}
	\frac{
        \Gamma^{2} \text{Re}
        \left[ \langle u_{\textbf{k}n}\vert \hat{v}_i \vert u_{\textbf{k}m}\rangle
        \langle u_{\mathbf{k}m} \vert \hat{v}_j \vert u_{\textbf{k}n}  \rangle \right ] 
	}
	{[\left(E_{F}-E_{\textbf{k}n})^{2}+\Gamma^{2}]([E_{F}-E_{\textbf{k}m}\right)^{2}+\Gamma^{2}]}\,, 
	\label{eq:Kubo_even} \\
	\sigma^{{\rm AHE}}_{ij} &= \frac{e^2\hbar}{2V\pi} \sum_{\textbf{k} n \neq m}
	\text{Im} \left[ \langle u_{\textbf{k}n}\vert \hat{v}_i \vert u_{\textbf{k}m}\rangle
	\langle u_{\textbf{k}m} \vert \hat{v}_j \vert u_{\textbf{k}n}  \rangle \right ] \label{eq:Kubo_odd} \\
	&\times \bigg\{ \frac{\Gamma ( E_{\textbf{k}m} - E_{\textbf{k}n} ) }{[\left(E_{F}-E_{\textbf{k}n})^{2}+\Gamma^{2}][E_{F}-E_{\textbf{k}m}\right)^{2}+\Gamma^{2}]} \notag \\
	&+ \frac{2\Gamma}{[\left(E_{\textbf{k}n}-E_{\textbf{k}m}][(E_{F}-E_{\textbf{k}m}\right)^{2}+\Gamma^{2}]} \notag \\
	&+ \frac{2}{( E_{\textbf{k}n} - E_{\textbf{k}m})^2} \text{Im}\ln 
	\frac{E_{\textbf{k}m} - E_F - i\Gamma}{E_{\textbf{k}n} - E_F - i\Gamma}
	\bigg\}. \notag
\end{align}
Here, $\sigma^{{\rm even}}$ denotes the time-reversal invariant component of the conductivity, which corresponds to the ordinary conductivity. $\sigma^{{\rm AHE}}_{ij}$ is the time-reversal odd part, which corresponds to the AHE. $E_{\textbf{k}n}$ and $u_{\textbf{k}n}$ respectively denote the Bloch energy and wavefunction for band $n$ and k-point $\textbf{k}$, $E_F$ denotes the Fermi level. The sum runs over all $\mathbf{k}$-points in the Brillouin zone. For practical evaluation, the sum is replaced by integral, which is evaluated by discretizing on a finite mesh.

In the $\Gamma \rightarrow 0$ limit, Eq. \eqref{eq:Kubo_even} becomes the well known Boltzmann constant relaxation time formula, which scales as $1/\Gamma$, whereas Eq. \eqref{eq:Kubo_odd} goes to the intrinsic formula, which is $\Gamma$ independent:

\begin{align}
   \sigma^{{\rm int}}_{ij}=  -
   \frac{2e{\hbar}}{V} \sum_{\textbf{k},m\neq n}^{\substack{ \\ n\ \text{occ.}\\ m\ \text{unocc.}}}
	\frac{\text{Im}\langle u_{\textbf{k}n}\vert \hat{v}_i \vert u_{\textbf{k}m}\rangle \langle u_{\textbf{k}m}\vert \hat{v}_j \vert u_{\textbf{k}n}  \rangle}{\left(E_{\textbf{k}n}-E_{\textbf{k}m}\right)^{2}} \,. 
	\label{eq:intrinsic}
\end{align}
We use the $\Gamma$ dependent formula since the intrinsic formula can be numerically difficult to evaluate and because including the $\Gamma$ broadening is more realistic since every system contains disorder in practice. For small $\Gamma$ our results will typically be very close to the intrinsic formula, however.

We have found that Eq. \eqref{eq:Kubo_odd} can in rare cases be numerically unstable and give wrong results, due to band degeneracies. These cases can be identified easily as then give the wrong symmetry and erratic $\Gamma$ dependence. In these situations we used a different Kubo formula, given by Eq. 5 in Ref. \cite{Li15}, which has the same $\Gamma \rightarrow 0$ limit and is more numerically stable.

For linear response calculation we use $\Gamma$ values $0.1 \uspace \text{meV}$, $0.5 \uspace \text{meV}$, $1 \uspace \text{meV}$, $5 \uspace \text{meV}$ and $10 \uspace \text{meV}$. We find that in most cases the AHE becomes $\Gamma$ independent for small $\Gamma$, which shows that the obtained small $\Gamma$ value is very close to the intrinsic value. We define a $\Gamma$ convergence parameter, which is defined as the relative AHE difference between the two lowest $\Gamma$ values. This parameter estimates how close is the lowest $\Gamma$ value to the intrinsic value. As shown in Fig. \ref{fig:convergence}(c) in most materials this parameter is less than few percent. Since the linear response calculation is very sensitive to the number of k-points we use a dense 250x250x250 k-mesh for the integration. To check the k-point convergence we also do a test calculation with 125x125x125 for most materials. As shown in Fig. \ref{fig:convergence}(d), the k-point convergence is strongly dependent on $\Gamma$: the smaller the $\Gamma$ is, the more k-points are needed. Nevertheless, even for $\Gamma = 0.1 \uspace \text{meV}$ the relative difference between AHE for 250x250x250 and 125x125x125 is smaller than $25\uspace\%$ for $85 \uspace \%$ of materials.

\begin{acknowledgments}
We  acknowledge  the  Grant  Agency  of  the  Czech  Republic  Grant  No.  19-18623Y, Ministry of Education of the
Czech Republic Grant LM2018110, EU FET Open RIA Grant No. 766566  and  support  from  the  Institute  of Physics  of  the  Czech  Academy  of  Sciences  and  the  Max Planck Society through the Max Planck Partner Group Programme. This work was supported by the Ministry of Education, Youth and Sports of the Czech Republic through the e-INFRA CZ (ID:90140). Y. Y. Acknowledges support from GP-Spin at Tohoku University
\end{acknowledgments}

\section*{Author Contributions}

J. \v{Z}, Y. S and Y. Z conceived and planned the project. J. \v{Z}, Y. Y and C.-G. O. have wrote the code for running and analyzing the calculations. J. \v{Z} has performed the calculations. All the authors have contributed to data analysis and the writing of the manuscript.

\section*{Competing Interests}

The authors declare no competing interests.

%\bibliography{refs,others}
%apsrev4-2.bst 2019-01-14 (MD) hand-edited version of apsrev4-1.bst
%Control: key (0)
%Control: author (8) initials jnrlst
%Control: editor formatted (1) identically to author
%Control: production of article title (0) allowed
%Control: page (0) single
%Control: year (1) truncated
%Control: production of eprint (0) enabled
%

\end{document}